\documentclass[conference]{IEEEtran}
\usepackage{cite}
\usepackage{amsmath,amssymb,amsfonts}
\usepackage{algorithmic}
\usepackage{graphicx}
\usepackage{textcomp}
\usepackage{xcolor}
\def\BibTeX{{\rm B\kern-.05em{\sc i\kern-.025em b}\kern-.08em
    T\kern-.1667em\lower.7ex\hbox{E}\kern-.125emX}}

\usepackage{microtype}
\usepackage{balance}
\usepackage[T1]{fontenc}
\usepackage{fancyvrb}
\usepackage{listings}
\usepackage{tikz}
\usetikzlibrary{arrows,automata, arrows.meta}

\usepackage[newfloat,frozencache,cachedir=./]{minted}

\usepackage{caption}

\newenvironment{code}{\captionsetup{type=listing}}{}
\SetupFloatingEnvironment{listing}{name=Listing}

\usepackage{csquotes}

\usepackage{url}
\definecolor{logging-blue}{RGB}{1, 130, 172}
\definecolor{dark-red}{rgb}{0.3,0.1,0.1}
\definecolor{dark-green}{rgb}{0.1,0.3,0.1}
\definecolor{dark-blue}{rgb}{0.1,0.1,0.5}
\usepackage[colorlinks,linkcolor=dark-red,citecolor=dark-green,urlcolor=dark-blue]{hyperref}
\usepackage[nameinlink]{cleveref}

\newcommand\copyrighttext{%
  \footnotesize \textcopyright 2021 IEEE. Personal use of this material is permitted. Permission from IEEE must be obtained for all other uses, in any current or future media, including reprinting/republishing this material for advertising or promotional purposes, creating new collective works, for resale or redistribution to servers or lists, or reuse of any copyrighted component of this work in other works. Cite this article as follows: J. Vykopal, P. Čeleda, P. Seda, V. Švábenský, and D. Tovarňák. \textit{Scalable Learning Environments for Teaching Cybersecurity Hands-on}, in Proceedings of the 51st IEEE Frontiers in Education Conference (FIE '21). Lincoln, Nebraska, USA, 2021. DOI: \href{https://doi.org/10.1109/FIE49875.2021.9637180}{10.1109/FIE49875.2021.9637180}.}
\newcommand\copyrightnotice{%
\begin{tikzpicture}[remember picture,overlay]
\node[anchor=south,yshift=12pt] at (current page.south) {\fbox{\parbox{\dimexpr\textwidth-\fboxsep-\fboxrule\relax}{\copyrighttext}}};
\end{tikzpicture}%
}

\begin{document}

\title{Scalable Learning Environments for Teaching Cybersecurity Hands-on}


\author{\IEEEauthorblockN{Jan Vykopal\IEEEauthorrefmark{1},
Pavel Čeleda\IEEEauthorrefmark{1}, Pavel Seda\IEEEauthorrefmark{1}, 
Valdemar Švábenský\IEEEauthorrefmark{1}, and Daniel Tovarňák\IEEEauthorrefmark{2}}
\IEEEauthorblockA{\IEEEauthorrefmark{1}Faculty of Informatics,
Masaryk University, Brno, Czech Republic\\
\IEEEauthorrefmark{2}Institute of Computer Science,
Masaryk University, Brno, Czech Republic\\
\{vykopal | celeda | seda | svabensky\}@fi.muni.cz, tovarnak@ics.muni.cz}}

\newcommand{\KYPO}{KYPO CRP}
\newcommand{\CSC}{CSC}

\maketitle
\copyrightnotice


\begin{abstract} 


This Innovative Practice full paper describes a technical innovation for scalable teaching of cybersecurity hands-on classes using interactive learning environments. Hands-on experience significantly improves the practical skills of learners. However, the preparation and delivery of hands-on classes usually do not scale. Teaching even small groups of students requires a substantial effort to prepare the class environment and practical assignments. Further issues are associated with teaching large classes, providing feedback, and analyzing learning gains. We present our research effort and practical experience in designing and using learning environments that scale up hands-on cybersecurity classes. The environments support virtual networks with full-fledged operating systems and devices that emulate real-world systems.

The classes are organized as simultaneous training sessions with cybersecurity assignments and learners' assessment. For big classes, with the goal of developing learners' skills and providing formative assessment, we run the environment locally, either in a computer lab or at learners' own desktops or laptops. For classes that exercise the developed skills and feature summative assessment, we use an on-premises cloud environment. Our approach is unique in supporting both types of deployment. The environment is described as code using open and standard formats, defining individual hosts and their networking, configuration of the hosts, and tasks that the students have to solve. The environment can be repeatedly created for different classes on a massive scale or for each student on-demand. Moreover, the approach enables learning analytics and educational data mining of learners' interactions with the environment. These analyses inform the instructor about the student's progress during the class and enable the learner to reflect on a finished training. Thanks to this, we can improve the student class experience and motivation for further learning.

Using the presented environments KYPO Cyber Range Platform and Cyber Sandbox Creator, we delivered the classes on-site or remotely for various target groups of learners (K-12, university students, and professional learners). The learners value the realistic nature of the environments that enable exercising theoretical concepts and tools. The instructors value time-efficiency when preparing and deploying the hands-on activities. Engineering and computing educators can freely use our software, which we have released under an open-source license. We also provide detailed documentation and exemplary hands-on training to help other educators adopt our teaching innovations and enable sharing of reusable components within the community.

\end{abstract}

\begin{IEEEkeywords}
cybersecurity education, interactive learning environment, sandbox, virtual machines, cyber range, educational data mining, learning analytics, learning technology 
\end{IEEEkeywords}

\section{Introduction}
\label{sec:intro}


Cybersecurity is a complex professional domain involving technology, people, information, and processes to enable assured operations~\cite{csec2017}. The dependence of modern society on digital technologies rapidly increased the demand for skilled workers. As a result, cybersecurity has become a focus area of K–12 and higher education, as well as professional learning. Educational institutions of all types expand their program and course offerings in cybersecurity for students and introduce re-skilling opportunities for mid-career professionals.

We support these efforts by designing and providing scalable environments for in-person or remote hands-on learning. These environments emulate real-world systems, applications, and infrastructures using virtual networks with full-fledged operating systems, devices, and applications used in authentic workplace settings. Learners' interaction with the emulated systems is driven by a serious game or a step-by-step tutorial facilitated by the learning environment with or without a human instructor's assistance. 
Regardless of the session's goal, the environments provide learning analytics. Students are informed about their progress during and after the session, which enables them to reflect on their learning. Instructors receive evidence of students' learning required for summative assessment, either a practical exam or a competition.

The environments can be deployed locally (either at individual hosts in a computer lab or at learners' desktops or laptops) or in a cloud. This flexibility enables many students to learn in a small environment or fewer students to learn in an extensive or complex environment. In both cases, students can practice from their school, workplace, home, or other places connected to the Internet. The environment can be repeatedly created for different classes on a massive scale or for each student on-demand.

From an instructor's perspective, the environments are described as code using open and standard formats, definitions of individual hosts and their networking, configuration of the hosts, and tasks that the students solve. Consequently, these components can be reused in other learning technologies.

Our environments have been iteratively improved and used in practice since 2013; the current, third generation is being developed since 2018. The learners reported they value the realistic nature of the environments that enable exercising theoretical concepts and tools in a safe way. The instructors stated that our approach saves their time required to prepare and deploy hands-on activities. We have released the presented interactive learning environments as two open-source projects~\cite{kypo-website,csc-website} so that other instructors can freely use them. We also provide detailed documentation and exemplary hands-on training to help others adopt our approach. The educators can thus focus on the topic and training itself and do not need to spend extra time setting up the learning environment.


This article is divided into seven sections. \Cref{sec:relatedwork} summarizes the state of the art of environments for teaching and learning cybersecurity skills. \Cref{sec:blocks} identifies building blocks of a cybersecurity hands-on class. \Cref{sec:approach} introduces reusable components of the building blocks, which we use to deliver our classes at scale. \Cref{sec:technologies} details the environments we have been developing using the described building blocks. \Cref{sec:experience} reports our experience from using these environments in teaching practice. \Cref{sec:conclusions} concludes the paper.

\section{Related Work} 
\label{sec:relatedwork}

Cybersecurity professionals require specialized education and hands-on training sessions to improve their skills and knowledge to secure the society. To enable the hands-on training sessions, cyber ranges and their lightweight alternatives are a valuable tool and a catalyst in these efforts. 

\subsection{Cyber Ranges}

Cyber ranges are interactive learning environments representing the network systems, tools, and applications. The main benefits of cyber ranges are providing (i) performance-based learning and assessment, (ii) real-time feedback, (iii) simulated on-the-job experience, (iv) an authentic environment where teams can collaborate to improve team capabilities~\cite{cyberrangesaguide2021,knupfer2020cyber}.

In the last decade, a number of commercial and non-commercial cyber ranges were developed, e.g., Cyber Range Instantiation System, DETER, ELTA Cyber Academy, Airbus Cyber Range, Silensec Cyber Range, and Circadence \cite{circadance2021,iaicrp2021,airbuscrp2021,silensec2021}. The comprehensive summary of cyber ranges is presented in \cite{ukwandu2020review,chouliaras2021cyber,Yamin2020}. These cyber ranges mainly differ in technologies, integrated capabilities, and functionalities.

Yamin et al.~\cite{Yamin2020} classify the functions and capabilities of cyber ranges into the following taxonomy (i) \emph{scenario}; (ii) \emph{monitoring}; (iii) \emph{scoring}; (iv) \emph{management}; and (v) \emph{teaming}. The scenario defines the execution environment and the execution steps. The monitoring includes the methods and tools supporting real-time data collection for further analysis. Scoring defines a way to evaluate the training participant's behavior. The management defines the roles and resources required in the system. Teaming includes an individual and a group of individuals with specific roles in the cybersecurity training (e.g., red teams or blue teams). This survey found that only a few publicly available cyber ranges and security testbeds support all of these functions. 

Pham et al. in~\cite{pham2016cyris} propose a Cyber Range Instantiation System for large-scale scenarios with a huge number of participants. They discuss that traditional approaches using dedicated and isolated physical computer infrastructure are inefficient. The main drawbacks of the traditional approach are expensive creation, maintenance, and low scalability in cases with tens, hundreds, or more participants. Being aware of this, they constructed virtual cyber range environments using the Kernel-based Virtual Machine (KVM) virtualization platform. They present their solution as unique in terms of security and other pre-installed features. However, they conclude their cyber range is not prepared for massive-scale environments that involve hundreds or thousands of virtual machines (VM).

Urias et al. in~\cite{urias2018cyber} discuss the limitations and needs for cyber range deployments. They highlight the rising difficulty of the simulation scenarios used in training sessions. Together with the expected functions and capabilities presented in \cite{Yamin2020}, scalability is one of the most critical requirements to consider when deploying cyber ranges \cite{urias2018cyber}.

Currently, many cyber ranges allow conducting training sessions with complex scenarios. However, they require sophisticated infrastructure, considerable resources, and workforce to prepare and maintain the cyber range before and during the training. Considering these, the cyber ranges are not suitable for less complex scenarios (e.g., to learn basic Unix commands in introductory university courses) with hundreds of students.




\subsection{Lightweight Labs}


Lightweight cybersecurity labs typically use virtual environments (e.g., VirtualBox) to run training scenarios on participants' personal computers. Using this approach, the instructors can organize training sessions with a huge number of participants without the need to provide and manage many centralized resources (such as in the cloud).

Du et al. in \cite{du2008seed} introduce SEED lab environment based on an instructional OS (Minix) and a production OS (Linux). The project developed over 30 labs covering a wide area of topics ranging from network security to security of mobile applications. They discuss that the students' feedback is important to improve the local-based environment. They identified two factors: \emph{efficiency} and \emph{effectiveness}. Efficiency focuses on the appropriateness and design of labs and effectiveness on student's learning output. Although hundreds of students can be served through this lab environment, the VM images are fixed. It is difficult for instructors to modify them, and thus it reduces the scalability.

Irvine et al.~\cite{irvine2017live,thompson2018individualizing,thompson2021labtainers} describe fully-provisioned Linux-based lab exercises with an emphasis on cybersecurity. They developed a framework consisting of 50 lab exercises named Labtainers. The solution is distributed as a single VM for either VirtualBox or VMware. The framework uses Docker containers to instantiate networked hosts within a single VM. Labtainers enable conducting training sessions with many participants since it requires fewer resources (hardware, software, workforce, and finances) compared to cyber ranges and dedicated VMs. Further, it can reduce the instructors' effort necessary to prepare the training since it does not require deploying all the scenarios and configurations on the server. The instructors can lack institutional IT equipment and staff to manage and deploy a fine-tuned lab environment. Although the provided solution applies network virtualization using Docker containers, this solution does not support functions and capabilities typical for cyber ranges presented in \cite{Yamin2020}. For instance, it does not allow creating teams or does not provide learning analytics and complex scoring visualizations.

To conclude the review of related work, cybersecurity hands-on education currently uses lightweight labs~\cite{thompson2021labtainers} or cyber ranges~\cite{ukwandu2020review, chouliaras2021cyber, Yamin2020}. The cyber ranges have its drawbacks in the difficult deployment and heavy resource requirements (despite being offloaded from students to instructors). The lightweight labs can serve a huge number of users using their local hosts. However, the labs do not provide all functions and capabilities of cyber ranges. To the best of our knowledge, there are no published attempts or open-source examples that benefit from both of these areas. 

%


\section{Building blocks of a Cybersecurity Hands-on~Class} \label{sec:blocks}

Cybersecurity hands-on classes can be held in many different formats. Students follow step-by-step instructions, solve a complex assignment autonomously, play a serious game (such as capture the flag or cyber defense exercise), or work together on a long-term project. Instructions and assignments can be provided by a tutor, which can be a human or a machine. Regardless of the format, students usually interact with a learning environment that features the following building blocks: \emph{sandbox}, \emph{class delivery method}, and \emph{learning analytics}. In this section, we define and detail these blocks.

\subsection{Sandbox}
Sandbox is an isolated environment for practicing cybersecurity skills. It is the essential component within the interactive learning environment. For the rest of the paper, we assume that the term \emph{sandbox} denotes either a single \emph{virtual machine} (VM), or a network of VMs. Users can run the sandboxes on their own hosts or access them remotely if deployed in a cloud.

VM is an instantiated representation of a \emph{virtual image} using a \emph{virtualization platform} and metadata defining hardware of the VM. \emph{Virtual image} is a snapshot of a computer disk that can be distributed as a file. The images typically carry the installation of an operating system with specific applications, tools, and files.

Once all the VMs are booted, networked, and running, i.e., they constitute a sandbox, users can interact with them. Two basic access methods can be used: \emph{console}\footnote{Also KVM, Keyboard, Video, and Mouse.} and \emph{remote access}. The former is provided by the virtualization platform and is always present, enabling users to interact with input and output devices of VMs. The latter relies on the installed operating system and its applications. The most common remote access protocol is Secure Shell (SSH) for Linux-based systems, and Remote Desktop Protocol (RDP) or PowerShell Remoting Protocol (PSRP) for Windows systems.

\subsection{Class Delivery Method}

The class delivery method is a computer-assisted instruction that employs the sandbox. It is provided as a part of the learning environment. It can use teacher- or student-centered approach. For instance, instruction can be provided using static (rich) text \emph{assignments} or \emph{serious games}. Some instructional methods require the assistance of an instructor, who also assesses students' progress or outcome. These methods include labs integrated with lectures, project-based learning, and problem-based learning~\cite{Sanders2017}. Other methods fully rely on a learning environment that presents assignments, provides feedback, or assesses students' outcomes without a human instructor's assistance. These methods include self-directed learning or automated tutoring systems~\cite{Sanders2017}.

\subsection{Learning Analytics}

Learning analytics (LA)~\cite{handbook-la2017} and educational data mining (EDM)~\cite{handbook-edm2010} are two growing disciplines that leverage data from educational contexts. They aim to understand and improve teaching and learning~\cite{hundhausen2017}. In this paper, we refer to components analyzing students' interactions in both the sandbox and the class delivery as \emph{learning analytics}. It can be used during and after the class with different goals. During the class, instructors can monitor students' progress and provide formative assessment to students. It can also be used for summative assessment, i.e., student testing and grading. After the class, LA enables students to reflect on the class and plan further learning. Instructors can use LA to improve the class for future runs.

\section{Scalable Approach} 
\label{sec:approach}

Preparing a hands-on class is a laborious and complex task. Therefore, instructors strive to use the above-mentioned building blocks repetitively. Less often, they also improve the class based on students' feedback or a technology change. The manual deployment of the sandbox and the class delivery is time-consuming, and it does not scale well for big classes.

This section describes reusable components of the building blocks introduced in \Cref{sec:blocks}. In particular, we introduce:
\begin{itemize}
    \item \emph{Topology definition} and \emph{provisioning definition}, two parts of \emph{sandbox definition}, which specifies the internal structure of the sandboxes (networks and hosts) and configuration of the hosts.
    \item \emph{Training definition} specifying the tasks and questions for students in the class.
    \item \emph{Learning analytics stack}, a key component for providing both formative and summative assessment to students.
\end{itemize}

These components make delivering cybersecurity hands-on classes scalable. The components are cornerstones of the two learning environments described in \Cref{sec:technologies}.

\subsection{Topology Definition}

\begin{figure}[!ht]
    \centering
    \includegraphics[width=0.35\textwidth]{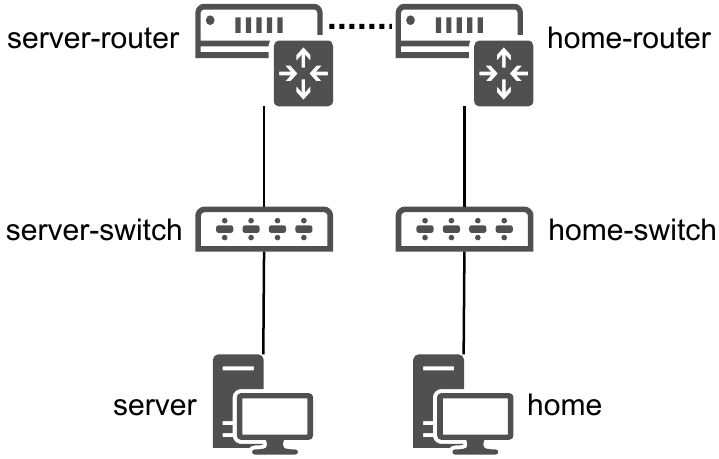}
    \caption{An example of a network topology of a sandbox.}
    \label{fig:topology_definition}
\end{figure}

Topology definition specifies 
a name of the definition, hosts, routers, networks, network mappings, router mappings, and host or router groups~\cite{topo-def}. \Cref{lst:sandbox-yaml} is an example of a topology definition of a sandbox shown in \Cref{fig:topology_definition} expressed in a YAML format. 


\begin{code}
\begin{minted}[fontsize=\footnotesize,numbers=left,xleftmargin=5mm]{yaml}
name: small-sandbox

hosts:
  - name: server
    base_box:
      image: debian-9-x86_64
      man_user: debian
    flavor: tiny1x2

  - name: home
    base_box:
      image: debian-9-x86_64
      man_user: debian
    flavor: tiny1x2

routers:
  - name: server-router
    cidr: 100.100.100.0/29
    base_box:
      image: debian-9-x86_64
      man_user: debian
    flavor: tiny1x2

  - name: home-router
    base_box:
      image: debian-9-x86_64
      man_user: debian
    cidr: 200.100.100.0/29
    flavor: tiny1x2

networks:
  - name: server-switch
    cidr: 10.10.20.0/24
    
  - name: home-switch
    cidr: 10.10.30.0/24

net_mappings:
    - host: server
      network: server-switch
      ip: 10.10.20.5

    - host: home
      network: home-switch
      ip: 10.10.30.5

router_mappings:
    - router: server-router
      network: server-switch
      ip: 10.10.20.1

    - router: home-router
      network: home-switch
      ip: 10.10.30.1

groups:
  - name: user-accessible
    nodes:
      - home
      - home-router
\end{minted}
\captionof{listing}{YAML topology definition of a sandbox.\newline}
\label{lst:sandbox-yaml}
\end{code}

This topology defines two hosts, two routers, two networks, and one named group. 
One host is named \texttt{server} and the other \texttt{home}. 
The definition specifies host properties, such as virtual image (\texttt{base\_box}) or the username for host provisioning (\texttt{man\_user}).
The hosts are assigned to respective networks with particular IP addresses (\texttt{net\_mappings}),
and the networks are assigned to the routers (\texttt{router\_mappings}). 
The definition also creates one named group (\texttt{user-accessible}), which will be used for provisioning.

\subsection{Provisioning Definition}
Provisioning definition specifies configuration changes at the hosts defined in the topology definition. It is used to customize the \emph{base boxes}, which contain only a minimal installation of a particular operating system (such as Ubuntu Server 20.04). The provisioning definition provides inputs for software configuration management system, which is responsible for installing and configuring the hosts deployed as base boxes. The provisioning definition contains prescriptions for the installation and configuration of applications. Also, it provides data specific to the particular class (such as installation of Apache web server and provisioning of a web application running at this server). 

The common approach is providing hosts as virtual images in binary files. However, these images cannot be easily updated or modified. In contrast, this approach gives instructors more flexibility in preparing the sandbox and allows them to reuse the parts of the definitions for another class.


Our approach uses Ansible for software configuration management. Ansible executes ordered lists of tasks provided as machine-readable YAML files~\cite{ansible}. \Cref{lst:ansible-yaml} is a simple example of two tasks installing a web server 
and provisioning files of a web application 
at the host named \texttt{server}. 


\begin{code}
\begin{minted}[fontsize=\footnotesize,numbers=left,xleftmargin=5mm]{yaml}
- hosts: server
  become: true
  tasks:
    - name: Install Apache, MySQL and PHP5
      apt:
        name: [apache2, mysql-server, 
               php5-mysql, php5]
        state: present
        update_cache: yes

    - name: Copy app to the web root
      copy:
        src: web-app/
        dest: /var/www/html
...
\end{minted}
\captionof{listing}{Ansible configuration file in YAML format.\newline}
\label{lst:ansible-yaml}
\end{code}


The provisioning definition can contain multiple files and directories following the Ansible conventions.

\subsection{Base Boxes} 

Base boxes can be provided by third parties or can be specifically tailored. Custom-made base boxes enable more flexibility, transparency, and better performance. However, their preparation is laborious and requires expert skills, which instructors may lack. For this reason, we provide open-source custom base boxes of common operating systems fine-tuned for use in our environments (both local and cloud).

The custom base boxes are created using Packer, an open-source tool for creating identical machine images for multiple platforms from a single source configuration~\cite{packer}.
Packer uses machine-readable configuration files, which define the process of creating the images from an ISO file containing the installation image or virtual hard drive.


\subsection{Training Definition}

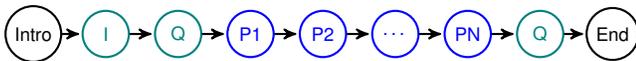
\begin{figure}[!ht]
    \centering
    \begin{tikzpicture}[
  ->,                  
  >=stealth',          
  shorten >=1pt,    
  auto,
  node distance=1.375cm,  
  thick,
  scale=0.7,
  every node/.style={scale=0.7}, 
  every node/.style={scale=0.7}, 
  font=\sffamily
  ]
  \node[state]   (A)                    {Intro};
  \node[state,color=teal] (I) [right of=A] {I};
  \node[state,color=teal] (B) [right of=I] {Q};
  \node[state,color=blue] (C) [right of=B]       {P1};
  \node[state,color=blue] (D) [right of=C]       {P2};
  \node[state,color=blue] (F) [right of=D]       {\dots};
  \node[state,color=blue] (G) [right of=F]       {PN};
  \node[state,color=teal] (H) [right of=G]       {Q};
  \node[state] (E) [right of=H]       {End};
  \path (A) edge node {} (I)
        (I) edge node {} (B)
        (B) edge node {} (C)
        (C) edge node {} (D)
        (D) edge node {} (F)
        (F) edge node {} (G)
        (G) edge node {} (H)
        (H) edge node {} (E);
\end{tikzpicture}
    \caption{Generic structure of training with several phases (P) and optional questionnaires (Q) and informative phases (I).}
    \label{fig:linear}
\end{figure}

Training definition specifies consecutive tasks that have to be solved by each student in the class. \Cref{fig:linear} shows a generic structure of a training. In each training phase (P), the student must complete a task and submit an answer (text) to prove the solution. Then, the student enters the next phase until completing the last phase. Each phase features one or more optional hints and a worked-out solution, which can be displayed if needed. Additionally, the training may include questionnaires (Q) at the beginning (pre-test) and the end (post-test), and information (I) for students, such as introductory instructions.

The training definition is a machine-readable description of the training and its training phases. \Cref{lst:training-definition} shows an example of a part of a training definition (metadata and the first training phase with three hints) expressed in a JSON format. 

\begin{code}
\begin{minted}[fontsize=\footnotesize,numbers=left,xleftmargin=5mm]{json}
{
  "title": "Secret Laboratory",
  "description": "A cybersecurity game.",
  "prerequisities": [ "Basic knowledge of Unix", 
                       "Basic networking"],
  "outcomes": ["nmap", "metasploit" ],
  "phases": [ {
    "title": "Introduction",
    "phase_type": "INFO",
    "order": 0,
    "estimated_duration": 0,
    "content": "Place for a background story."
}, {
    "title": "Looking for a vulnerable service.",
    "max_score": 100,
    "level_type": "TRAINING",
    "order": 1,
    "estimated_duration": 5,
    "flag": "service-name-1.23",
    "content": "Now you need to scan the server 
     to find possible vulnerabilities. The IP 
     address of the server is **10.1.26.9** . 
     The name of the vulnerable service starts 
     with \"s\". \n\n
     As a flag, submit the name of 
     the vulnerable service in the following 
     format: _service-version_. All characters 
     are lowercase. For example: _dvwa-2.050_.",
    "solution": "```root@attacker:~# nmap -sV 
     -p 10000 10.1.26.9\n```\n\n
     The flag is: **service-name-1.23**",
    "hints": [ {
      "title": "Which port should you scan?",
      "content": "The vulnerable service is 
       running on port 10000. You can also pass 
       this information to nmap 
       (**-p \"port range\"**).",
      "hint_penalty": 10,
      "order": 1
}, {
      "title": "Which option gives you the 
                 version?",
      "content": "To determine the name and 
       version of the service, you need to pass 
       the argument **-sV** to nmap - 
       service/version detection.",
      "hint_penalty": 15,
      "order": 2
}, {
      "title": "Which  tool should you use?",
      "content": "You should use **nmap** to 
       scan the server (see **man nmap**).",
      "hint_penalty": 10,
      "order": 0
    } ],
    "incorrect_flag_limit": 100,
}, {
    "...": "..."
}
\end{minted}
\captionof{listing}{Training definition in a JSON format.\newline}
\label{lst:training-definition}
\end{code}

Structuring the training to linear phases guides the students by dividing the complex assignment into smaller, more digestible units. In addition, the predefined order of tasks (levels) enables to practice processes or phases in the correct order (e.g., typical cybersecurity attack lifecycle phases). This is a different format than the popular \textit{jeopardy} game~\cite{davis2014}, where players can choose any task from available tasks at any time.

\subsection{Learning Analytics Stack}

Learning analytics stack is a mechanism of processing events from the learning environment. The events capture interactions of the student with the environment, particularly sandbox and class delivery method (training). The events can capture various actions of the student. These can include: starting or finishing the particular training or its individual phase; submitting a correct or incorrect answer in a training phase; displaying a hint or solution of the phase; logging into a host in a sandbox; typing a command at the command line of a host in the sandbox.

\newcommand{\cmdpart}[2]{\underbrace{\texttt{#1}}_{#2}\ \ \ }
\begin{figure*}[ht]
\[
\setlength{\jot}{7pt} 
\footnotesize
\begin{split}
&\cmdpart{Feb 17 2021 9:17:33}{timestamp}
\cmdpart{username=\textquotedbl root\textquotedbl}{username}
\cmdpart{client}{hostname}
\cmdpart{src=\textquotedbl 10.10.40.5\textquotedbl}{host\ IP\ address} \\
&\hspace*{6mm}
\cmdpart{wd=\textquotedbl /home\textquotedbl}{working\ directory}
\cmdpart{cmd=\textquotedbl ssh alice@server\textquotedbl}{command}
\cmdpart{cmd\_type=\textquotedbl bash\textquotedbl}{command\ type}
\cmdpart{uid=\textquotedbl 1\textquotedbl}{sandbox\ ID}
\end{split}
\]
\caption{A log entry for a command executed on one machine in a sandbox.}
\label{fig:command-local}
\end{figure*}

The events are machine-readable strings logged using the standard Syslog protocol~\cite{rfc5424} in a predefined format, i.e., they can be produced by any application capable of Syslog logging. The learning environment forwards to and stores all events at the central storage, which transforms and further processes them. The central storage uses the ELK stack (Elasticsearch, Logstash, and Kibana)~\cite{elkstack}, which provides interfaces for both learning analytics applications and instructors.

For example, \Cref{fig:command-local} and \Cref{lst:command-json} show how a command \verb!ssh alice@server! executed by a student in the Linux terminal at a machine in the sandbox is timestamped and logged into Syslog as a string (\Cref{fig:command-local}). Then, it is transferred and stored at the central storage as an entry for further processing (\Cref{lst:command-json}).

\begin{code}
\begin{minted}[fontsize=\footnotesize,numbers=left,xleftmargin=5mm]{json}
{
    "timestamp": "2021-02-17T09:17:33+02:00",
    "username": "root",
    "hostname": "client",
    "host_ip": "10.10.40.5",
    "wd": "/home",
    "cmd": "ssh alice@server",
    "cmd_type": "bash-command",
    "sandbox_id": "1"
}
\end{minted}
\captionof{listing}{An entry at the central storage created from the Syslog record in \Cref{fig:command-local}.\newline}
\label{lst:command-json}
\end{code}

Another example is an event of submission of an incorrect answer (flag in a capture the flag game) \texttt{.invoices2019} to a task in a training phase (a level of the game). This event is timestamped and logged into Syslog (\Cref{lst:flag-local}) and finally arrives at the central storage. 

\begin{code}
\begin{minted}[fontsize=\footnotesize,numbers=left,xleftmargin=5mm]{json}
{
    "flag_content": ".invoices2019",
    "actual_score_in_level": 100,
    "total_score": 300,
    "game_time": 3045985,
    "timestamp": 1610618680221,
    "type": "events.trainings.WrongFlagSubmitted",
    "count": 1,
    "user_ref_id": 19,
    "phase_id": 36,
    "training_run_id": 28,
    "training_instance_id": 12,
    "training_definition_id": 7,
    "sandbox_id": 104,
    "pool_id": 40
}
\end{minted}
\captionof{listing}{An entry created by the training portal.\newline}
\label{lst:flag-local}
\end{code}



\section{Learning environments} \label{sec:technologies}

This section introduces two learning environments we developed based on the common components described in \Cref{sec:approach}. KYPO Cyber Range Platform (KYPO CRP)~\cite{kypo-website} is a cloud-based platform designed for running multiple classes in parallel or classes requiring sandboxes with many hosts. Cyber Sandbox Creator (CSC)~\cite{csc-website} is a lightweight, distributed lab environment using a commercial off-the-shelf computer in the lab or students' own desktop or laptop. 

Both environments use the same formats for topology, provisioning, and training definitions, and same formats of events processed by learning analytics stack. The key difference is the used virtualization technology for sandbox instantiation. This is most evident for base boxes, which are almost identical except for the features and limits bound to the different underlying virtualization technologies. Another important difference lies in user roles and access control to sandbox and training instances. \Cref{tab:kypo-csc} highlights the differences between these two environments.

\begin{table*}[!ht]
\caption{The comparison of features and capabilities of KYPO Cyber Range Platform and Cyber Sandbox Creator}
\label{tab:kypo-csc}
\setlength{\tabcolsep}{4pt}
\centering
\begin{tabular}{lcc}
\textbf{Feature} & \textbf{\KYPO} & \textbf{\CSC}  \\
\hline
Required hardware & Cloud of tens of servers & Desktop/laptop \\
Virtualization technology  & OpenStack & VirtualBox \\
Deployment & Advanced user & General user \\
Max. num. of hosts in a sandbox for each student & Up to free cloud resources & $\approx$ 5 \\
Number of parallel classes & Up to free cloud resources & 1  \\
Preparation effort at the students' side & None & Low \\
User access  & Remote (web browser, SSH, RDP) & Local \\
Sandbox definition visible to students & No & Yes \\
Task presentation & Integrated & Separate \\
\end{tabular}
\end{table*}

Both environments have been released as open-source software with documentation and examples of training.

\subsection{Cloud-based Learning Environment}

\begin{figure}[!ht]
    \centering
    \includegraphics[width=\columnwidth]{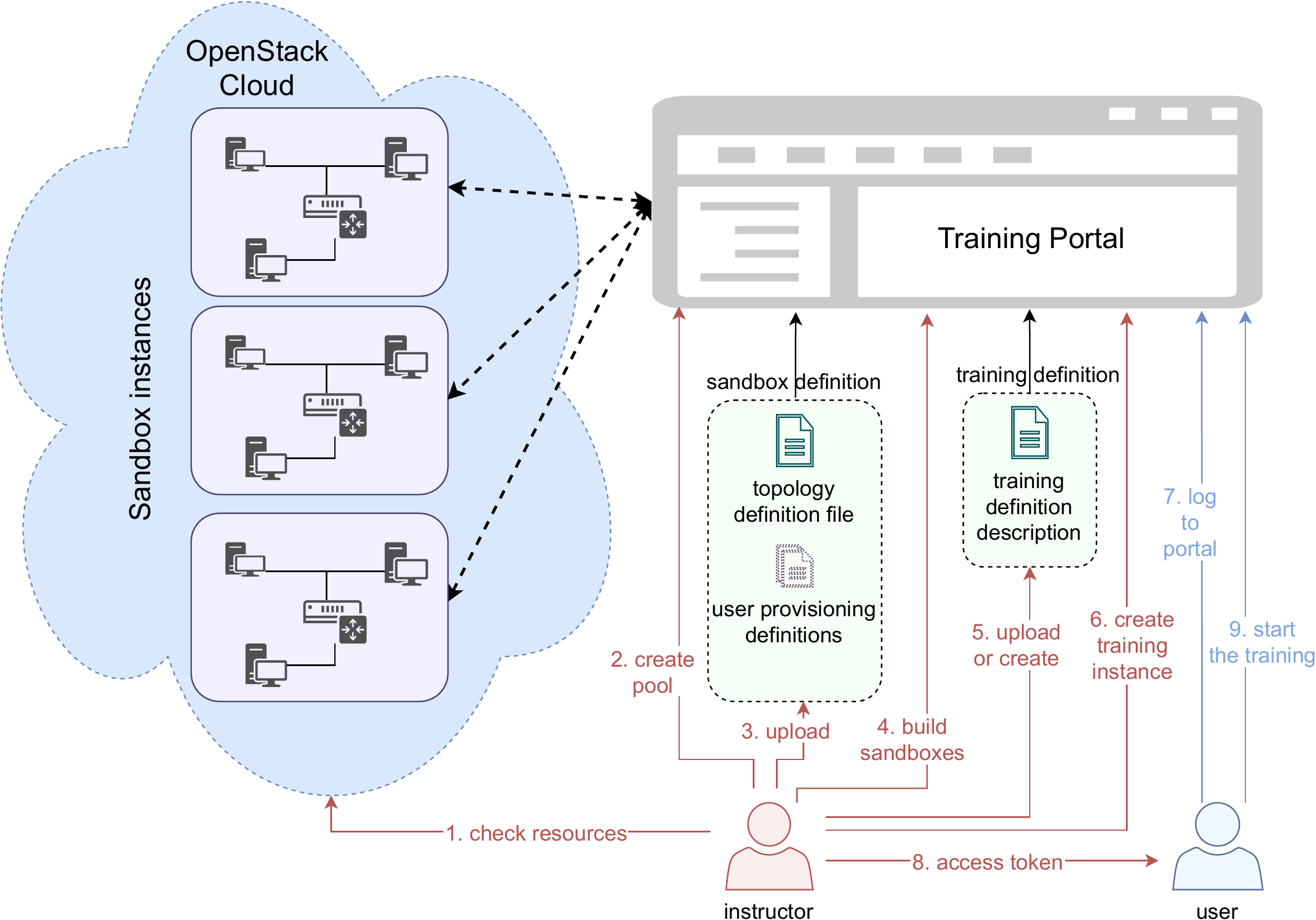}
    \caption{The principle of how the learning environment \KYPO{} works.}
    \label{fig:KYPOCRP}
\end{figure}

The design of the learning environment \KYPO{} is shown in \Cref{fig:KYPOCRP}.
First, the instructor checks that estimated resources for the training session are available in the cloud. The estimated resources are based on the number of students in the class and the size of the sandbox. After that, the instructor logs in to the web interface of \KYPO{} and allocates a pool of sandboxes for the class. The pool size is usually set to the number of students plus a few more for a reserve. Further, the instructor provides a Gitlab repository with sandbox and provisioning definitions. The repository may enable auxiliary services of the learning analytics stack. \KYPO{} then builds all sandboxes in the allocation pool in the cloud using sandbox and provisioning definitions. 

Next, the instructor creates or imports a training definition describing the task assignments. As the last step, the instructor creates a training instance for a particular class. The instance defines the start and end times of the training session and a respective pool of sandboxes. \KYPO{} thus enables instructors to run multiple (different) classes in parallel.

\begin{figure*}[!ht]
    \centering
    \includegraphics[width=\linewidth]{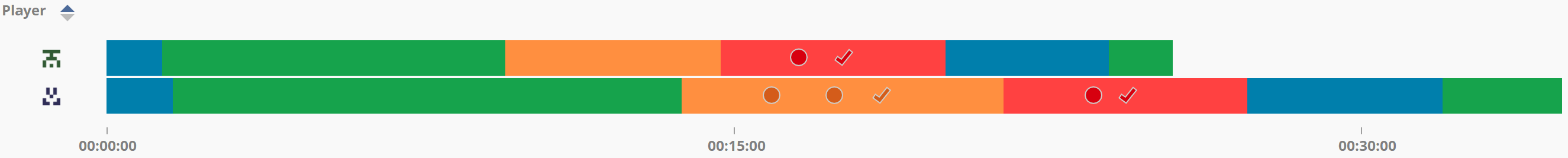}
    \caption{A part of the instructor's dashboard showing progress of two students working in \KYPO{}. Colored bars represent different completed training phases. A dot means a hint displayed by a student, and a tick solution displayed. At first glance, the instructor can see the first player performs better than the second one who may need additional help. Here, we present a view with students' avatars instead of their real names.}
    \label{fig:kypo-instructor}
\end{figure*}

The student logs into the web interface of \KYPO{} (training portal) and starts the training by entering an \emph{access token} provided by the instructor. If the entered access token is correct, an available sandbox instance from the pool is assigned to the particular student. The student then starts solving the tasks defined in the training definition by interacting with the sandbox instance hosted in the cloud. \KYPO{} provides access to the sandbox host using a web browser, presents tasks, provides predefined on-demand hints, checks submitted answers, and collects events from its web interface and allocated sandboxes. The instructor can monitor the progress of all students in the class during the training at a dashboard as shown in \Cref{fig:kypo-instructor}. When the training is over, analyses of student progress are available both to the student and instructor. 


\subsection{Lightweight Learning Environment}

\begin{figure}[!ht]
    \centering
    \includegraphics[width=0.45\textwidth]{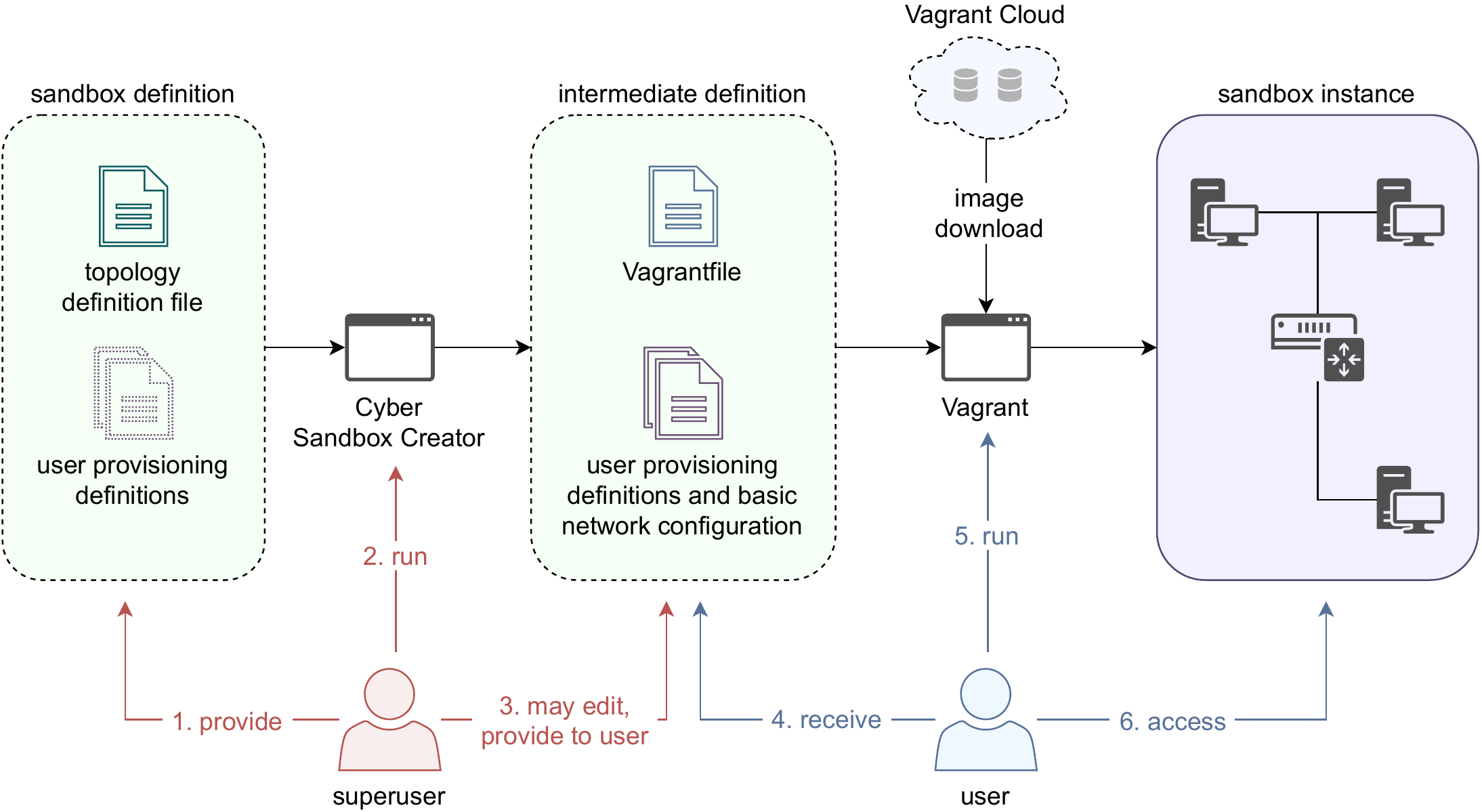}
    \caption{The principle of how the learning environment \CSC{} works.}
    \label{fig:CSC}
\end{figure}

The design of the learning environment \CSC{} is shown in \Cref{fig:CSC}. First, a superuser (instructor) writes or reuses a sandbox definition, from which an intermediate definition is generated. This can be further extended by provisioning definition, which specifies the software and configuration of the machines in the sandbox.

The user (student) receives this intermediate definition and builds the sandbox locally using a single command. This is achieved by using Vagrant, which downloads images of operating systems for the VMs (base boxes), and provisions them according to the superuser's configuration. As a result, the sandbox instance is ready on the student's computer.

The student performs the hands-on tasks in the sandbox instance. The task assignments and scaffolding are delivered via separate software, such as the training portal from the learning environment \KYPO{} or CTFd~\cite{chung2017}. Moreover, the superuser can enable auxiliary services in the sandbox definition, such as command logging from the learning analytics stack. As a result, the commands submitted by the student in the sandbox are forwarded to a central storage, where they can be processed further or viewed by an instructor.

\subsection{Use Cases}

Since both environments are based on common components, they enable the following use cases:

\subsubsection{Running the same training in either environment}
The common formats of training and sandbox definitions enable running training sessions in either \KYPO{} or \CSC{} without additional instructor's effort. The instructor can consider which features of the particular environment better fulfill the needs of a particular class (see \Cref{tab:kypo-csc}) and choose a more suitable environment. An example training that demonstrates this feature is publicly available at~\cite{junior-hacker}.

\subsubsection{Sharing of created training between organizations}
The human-readable formats of training and sandbox definitions enable their use in other organizations in the \KYPO{} and \CSC{} learning environments. Alternatively, it simplifies the adoption in different environments if they are based on the same or similar building blocks. For instance, training definitions could be easily transformed to a learning environment that is based on different technology for sandbox provisioning. Or, a sandbox definition of a typical small enterprise network could be used for different training sessions.


\section{Teaching Experience}
\label{sec:experience}

To demonstrate the usability of the presented learning environments for teaching cybersecurity, we now describe recent teaching experience. Since the year 2019, about 380 students used our learning environments in 31 training sessions. 
The students were undergraduates and graduates, professional learners, and selected high school students attending a cybersecurity competition. While most students received the training, some students were involved in creating the training and cybersecurity games as described in~\cite{svabensky2018kypolab}. They were able to create sandbox and training definitions using the learning environment \CSC{} and run the training for their peers in either the learning environment \KYPO{} or \CSC{}.

Some sessions were held by other instructors at other institutions in several European countries. Both students and instructors were able to run the training in the environment \CSC{} without our in-class assistance.

The learning environment \KYPO{} was successfully deployed in a private cloud at our university, Brno University of Technology, Czech Republic, and Swedish Defence Research Agency (FOI).

\subsection{Classes Aimed at Formative Assessment}

When teaching practical lab sessions at university courses, our goal is that students gain hands-on experience with using various cybersecurity tools. In this low-stakes context, we do not care if the students see the sandbox definition, so we usually decide to use \CSC{}. Regardless of the number of students, everyone deploys the sandbox locally on their own computer. This way, we have taught classes from 10 up to 200 students.

However, the teachers need to allocate time for preparing detailed setup instructions, as well as be ready to troubleshoot the setup. The students use a wide range of host operating systems (e.g., Windows, MacOS, and different Linux distributions), and the hardware configuration of their devices varies widely. As a result, some students may experience early difficulties before the environment is ready to run on their machines. In particular, a few students encountered issues with the installation of VirtualBox.

Alternatively, we may opt in for the \KYPO{}. The setup is simpler for students, as the instructors prepare sandboxes in the cloud, and students access them remotely using a web browser or SSH. However, the number of sandboxes we can host is limited by the cloud's resources. We usually teach classes from 10 to 30 students with this setup. An additional risk is that if the cloud space is shared, external users running other computational tasks in the cloud may jeopardize the stability of the environment and user experience. This risk is not associated with the \CSC{}, because if the students experience low responsiveness, they only need to close other unnecessary applications running on their computers.

Regardless of the environment used, we can collect command histories of students solving the tasks~\cite{my-2021-FIE-logging} and provide them with formative feedback. This includes explaining what they did well and what they can improve, for example, how to address frequently occurring mistakes.

\subsection{Classes Aimed at Summative Assessment}

When we need to perform summative assessment, such as during a final exam or a competition, we need to hide the sandbox definitions from students. Therefore, we only use the \KYPO{} environment for this use case. We can control the visibility of hosts in the sandbox topology so that students are initially aware only of a limited number of machines. They also cannot directly see what applications are running there, and so the setup mimics more realistic situations. 

\section{Conclusions}
\label{sec:conclusions}

Cybersecurity hands-on classes take place in a laboratory or an interactive learning environment featuring real-world tools, systems, and applications. The preparation of the classes requires not only pedagogical skills of choosing suitable instructional methods for the class but also proficiency in software development, deployment, and IT operations. 

In this paper, we demonstrated a technical innovation for enhancing cybersecurity classes. We support cybersecurity educators by providing scalable and reusable building blocks of the classes. These range from open formats for the description of technical environment, through content, to ready-made artifacts applicable in various contexts.

We exploit these blocks while developing two learning environments, \KYPO{} and \CSC{}. Each environment has its benefits and limitations, which determine its suitability for a particular use case. While \KYPO{} is more appropriate for summative assessment and classes requiring an extensive or complex networked environment, it relies on a dedicated cloud and personnel who can manage it. In contrast, \CSC{} is more suitable for big classes featuring a small network environment that can be deployed at students' hosts. Regardless, classes deployed in one environment can be easily deployed in the other if needed. In addition, open definitions of formats enable educators to enhance and edit the existing lab environments without much additional effort. Both environments \KYPO{}~\cite{kypo-website} and \CSC{}~\cite{csc-website} have been released as open-source software so educators can freely use them in their teaching practice.


\section*{Acknowledgment}
This research was supported by the Security Research Programme of the Czech Republic 2015--2022 (BV III/1--VS) granted by the Ministry of the Interior of the Czech Republic under No. VI20202022158 -- Research of New Technologies to Increase the Capabilities of Cybersecurity Experts.


\bibliographystyle{IEEEtran}
\bibliography{references}


\end{document}